\documentclass[10pt,twocolumn,a4paper,fleqn]{article}

\usepackage[numbers,sectionbib,square]{natbib} 

\usepackage{balance}
\newcommand{\myemail}{{\let\thefootnote\relax\footnote{$\star$
    \texttt{Lukasz.Bratek@pk.edu.pl}}}}
\newcommand{\mytitle}[1]{\begin{quotation}{\bf\huge\noindent #1}
    \end{quotation}}
\newcommand{\myabstract}[1]{\begin{quotation}\noindent{\bf Abstract.}{
    \small #1}\end{quotation}}
\newcommand{\mykeywords}[1]{\begin{quotation}\noindent{\bf Keywords:}{
    \small #1}\end{quotation}}    

\usepackage[utf8]{inputenc}
\usepackage[T1]{fontenc}

\usepackage{graphicx}

\usepackage{amsmath,amssymb}
\usepackage[scriptsize,bf]{caption}
\usepackage{float}
\usepackage{placeins}
\usepackage{xspace}

\newcommand{\eV}{\,{\rm eV}}

\newcommand{\sqm}{SQM\xspace}
\newcommand{\msun}{{M}_{\odot}}
\newcommand{\mearth}{{M}_{\oplus}}
\newcommand{\erad}{E_\mathrm{T}}
\newcommand{\nb}{N_{\!\!\:\;\!\!{B}}}
\newcommand{\ampl}{x}

\newcommand{\km}{\mathrm{km}}
\newcommand{\keV}{\mathrm{keV}}
\newcommand{\MeV}{\mathrm{MeV}}

\newcommand{\erg}{\mathrm{erg}}

\newcommand{\m}{\mathrm{m}}

\newcommand{\epsat}{\epsilon_{\!\!\;s}} 
\newcommand{\excen}{{\epsilon}_{\mathrm{exc}}}
\newcommand{\meanexcen}{\bar{\epsilon}_{\mathrm{exc}}}

\begin{document}
\twocolumn[\begin{@twocolumnfalse}

\bigskip\bigskip\bigskip

\mytitle{Oscillations of hypothetical strange stars 
as an efficient source ultra-high-energy particles}

\medskip

\begin{center}
{\large  Joanna Ja{\l}ocha$^{*}$, {\L}ukasz Bratek}
\\
\medskip
\begin{tabular}{@{}l}
{\small Department of Physics, Cracow University of Technology,  
Podchor\k{a}{\.z}ych 1, PL-30084 Krak{\'o}w, Poland}
\end{tabular}\\
\bigskip \texttt{Preprint v1: proc. 7th International Symposium on Ultra High Energy Cosmic Rays (UHECR2024)
 17-21 November 2024, Malargüe, Mendoza, Argentina\\}
\end{center}


\myabstract{
We investigate the dynamical behavior of strange quark matter (SQM) objects, such as stars and planets, 
when subjected to radial oscillations induced by tidal interactions in stellar systems. 
Our study demonstrates that SQM objects can efficiently convert mechanical energy into hadronic energy
 due to the critical mass density at their surfaces of \( 4.7 {\times} 10^{14} \, \mathrm{g\,cm}^{-3} \),
  below which \sqm becomes unstable and decays into photons, hadrons, and leptons. We show that even small-amplitude radial oscillations, 
  with a radius change of as little as 0.1\%, can result in significant excitation energies near the surface of SQM stars. 
  This excitation energy is rapidly converted into electromagnetic energy over short timescales (approximately \(1\,\mathrm{ms}\)),
   potentially leading to observable astrophysical phenomena. 
    Higher amplitude oscillations may cause fragmentation or 
   dissolution of \sqm stars, which has important implications for the evolution of binary systems containing \sqm objects and the 
   emission of gravitational waves.}
   
\mykeywords{SQM, strange quark stars, tidal interactions, radial oscillations, energy conversion
}

\renewcommand{\thefootnote}{\fnsymbol{footnote}}  
\bigskip
\end{@twocolumnfalse}]
\renewcommand{\thefootnote}{\fnsymbol{footnote}}  
\footnotetext[1]{\texttt{Joanna.Jalocha-Bratek@pk.edu.pl}}
\setcounter{footnote}{0}                          
\renewcommand{\thefootnote}{\arabic{footnote}}    


\section{Introduction}
\label{sec:intro}

Strange quark matter (SQM) has been hypothesized to be the true ground state of baryonic matter since the seminal 
work by Witten \cite{1984PhRvD..30..272W}. Neutron stars composed of \sqm were studied soon after \cite{1986ApJ...310..261A}. 
Under extreme conditions of pressure and density, matter containing up, down, and strange quarks can be more stable than 
ordinary nuclear matter.

Recent observations have renewed interest in possible SQM objects, especially of planetary mass. \sqm could exist at any mass down to planetary values \cite{1996astro.ph..4035W}, and even in smaller “strange nuggets” \cite{1984PhRvD..30.2379F}. Pulsar systems such as PSR B1257+12 show planets with unusually high densities \cite{2017ApJ...848..115H,Kuerban_2020}, hinting these objects may be \sqm.

In stellar or planetary systems, tidal interactions are common, inducing oscillations in compact objects. Predicting the signatures and stability of such oscillations in SQM matter is essential for astrophysical modeling. Previous work explored \sqm stars and their gravitational-wave properties \cite{2015ApJ...804...21G}, but the specific role of small radial oscillations remains less investigated.

Guided by these considerations, we study the behavior of \sqm stars and planets undergoing radial oscillations \cite{Kutschera_2020}. Our main goal is to see how mechanical energy near the surface translates into hadronic and electromagnetic energy, given that \sqm becomes unstable below a critical density \( \rho_0 \sim 4.7 \times 10^{14} \,\mathrm{g\,cm}^{-3} \). Even a small (\(\sim0.1\%\)) change in radius can push the surface layer below this threshold density, triggering rapid energy release. Such processes could have observable effects, especially in binary systems or gravitational wave sources.

\section{Objectives and Key Concepts}
\label{sec:objectives}

\subsection*{Main Objectives}
\begin{itemize}
\item Investigate the dynamical behavior and stability of \sqm stars or planets subjected to radial oscillations due to tidal forces.
\item Understand how mechanical (oscillatory) energy is converted into hadronic and electromagnetic energy, focusing on surface instabilities.
\item Assess astrophysical consequences, including fragmentation of \sqm objects, the potential for gravitational wave emission, and other observational effects.
\end{itemize}

\subsection*{Strange Quark Matter (SQM)}
SQM is a hypothetical form of matter with up, down, and strange quarks in roughly equal proportions. Under the MIT bag model \cite{1984PhRvD..30..272W}, \sqm can be more stable than nuclear matter at high densities, possibly forming large or small (nugget) objects.

\subsection*{Saturation Density \(\rho_0\)}
Below a critical density \(\rho_0\), \sqm decays into photons, hadrons, and leptons. In our model,
\[
\rho_0 = 4.665 \times 10^{14} \,\mathrm{g\,cm}^{-3}.
\]
For densities above \(\rho_0\), quark interactions stabilize \sqm in the MIT bag model.

\subsection*{Radial Oscillations}
Monopole or breathing-mode oscillations are uniform expansions/contractions of a star or planet. Even minor radial changes can alter the near-surface density and stability \cite{Kutschera_2020}.

\section{Methodology}
\label{sec:method}

We examine SQM star or planet evolution under tidal interactions in binary systems, incorporating excitation and energy conversion processes into stellar dynamics. Our approach includes:

\begin{itemize}
\item Modeling radial oscillations and their damping via electromagnetic energy release.
\item Analyzing maximum oscillation amplitudes for which SQM objects remain intact.
\end{itemize}

\subsection*{Numerical Simulations}
We solved Tolman-Oppenheimer-Volkoff (TOV) equations with equations of state for \sqm. Radial oscillations were induced by perturbing equilibrium configurations and following changes in density/pressure. 
The resulting excitation energy and potential for fragmentation were evaluated.

\subsection*{Stability Analysis}
We sought the maximum oscillation amplitude before fragmentation or dissolution. Tidal forcing inputs energy; electromagnetic release drains it. If release is fast enough, \sqm may remain stable unless amplitudes grow too large.

\subsection*{MIT Bag Model Details}
\label{sec:bag_model}

We adopt the MIT bag model \cite{PhysRevD.9.3471,1984PhRvD..30..272W}, in which:
\[
\rho c^2 = \varepsilon_q + B,
\]
with \(B\) the bag constant and \(\varepsilon_q\) the fermion energy density. Electrons and a finite s-quark mass also enter the equation of state \cite{bratek2024}. We set \( B=60\,\mathrm{MeV\,fm}^{-3}\), \( m_s=150\,\mathrm{MeV}\). Under these choices, \sqm is stable at zero pressure.

\section{Results}
\label{sec:results}

\begin{figure}
\centering
\includegraphics[width=0.5\textwidth]{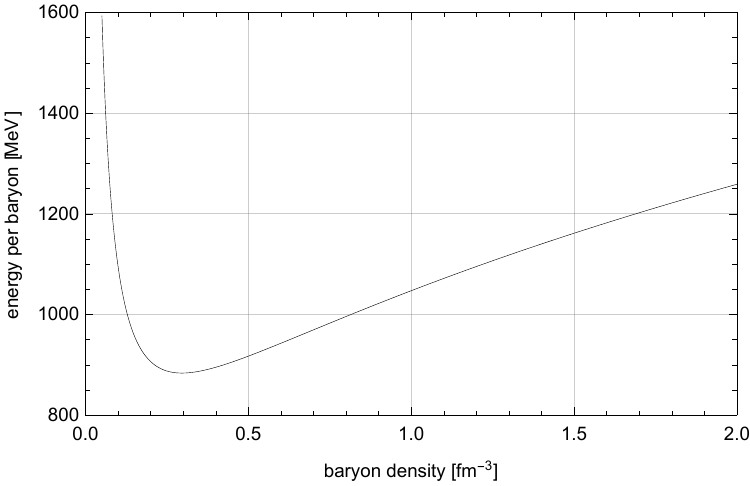}
\caption{\label{fig:energy_per_baryon}Energy per baryon vs.\ baryon density for \sqm in the MIT bag model with \( B=60\,\mathrm{MeV\,fm}^{-3}\), \(m_s=150\,\mathrm{MeV}\). The minimum at \(n_0\) identifies the saturation density.}
\end{figure}

Using chemical equilibrium and charge neutrality in the bag model, we derive the energy per baryon \(\varepsilon/n\) as a function of baryon density \(n\). Figure \ref{fig:energy_per_baryon} shows the minimum at saturation density \(n_0\), where \sqm is most stable.

\subsection*{Excitation Energy per Baryon}
Let \(\epsilon_{\text{exc}}\) be the energy above the minimum from density fluctuations near \(n_0\). Expanding around \(n_0\),
\[
\epsilon_{\text{exc}} = \tfrac12 \left.\tfrac{d^2\epsilon}{dn^2}\right|_{n_0}(\Delta n)^2,
\]
where \(\Delta n=n(x)-n_0\). The convexity parameter \(\beta_s\) is
\[
\beta_s = \frac{n_0^2}{\epsilon_{\text{min}}}\,\left.\frac{d^2\epsilon}{dn^2}\right|_{n_0},\quad \epsilon_{\text{min}}=\epsilon(n_0).
\]
A radial fraction \(x=\Delta R/R\) measures the amplitude of monopole oscillations. Integrating \(\epsilon_{\text{exc}}(r)\) over volume gives the total excitation \(E_{\text{exc}}\).

\section*{Scaling Laws for Energy in Monopole Perturbations}
\label{sec:scaling}

At the saturation point \(n_s=0.296136\,\mathrm{fm}^{-3}\), \(\epsat=883.623\,\mathrm{MeV}\), \(\beta_s=0.307124\). The excess \(\Delta\epsilon\) from small density changes \(\Delta n\) around \(n_s\) reads \(\Delta\epsilon=\tfrac12\,\epsilon''(n_s)(\Delta n)^2\).

\noindent\textbf{Small-mass \sqm objects (nearly constant density).}  
When \(n(r)\approx n_s\), and \(r\to r(1+x)\) for \(x\ll1\), the density changes by \(\Delta n=3n_s\,x\). Then
\[
\Delta\epsilon=\tfrac92\,\beta_s\,\epsat\,x^2\approx1221.22\,x^2\,\mathrm{MeV}.
\]
Multiplying by baryon number \(\nb=Mc^2/\epsat\approx3.794\times10^{51}\,(M/\mearth)\) yields the total energy to radiate:
\begin{equation}\label{eq:eradA}
\erad=\tfrac92\,\beta_s\,M c^2\,x^2\approx7.416\times10^{48}\,\mathrm{erg}\,\tfrac{M}{\mearth}\,x^2.
\end{equation}
This scales as \(x^2\). E.g., for an Earth-mass \sqm planet with \(x=10^{-5}\), \(\erad=7.4\times10^{38}\,\mathrm{erg}\).

\noindent\textbf{Higher-mass \sqm stars.}  
Keeping \(x\) fixed, only a thin subsaturation zone near the surface contributes to the excitation. By expanding solutions to Einstein’s equations, we find to leading order:
\begin{equation}\label{eq:eradB}
\begin{aligned}
&\erad=18\pi\,\beta_s^2\,\epsat\,n_s\,\frac{c^2\,R^4}{GM}\,H_{\Phi}\,x^3,\\
&\hspace{0.1\textwidth}\frac{\delta R_\ampl}{R}\equiv\frac{3\beta_s}{\Phi}\Bigl(1-\frac{7}{3}\Phi\Bigr)x\ll1,
\end{aligned}
\end{equation}
where \(\Phi=GM/(Rc^2)\) and \(H_\Phi\) is a relativistic correction. This scales as \(x^3\). For a $1.4\,\msun$ star at $R=10.31\,\km$, $\erad\approx6.634\times10^{54}\,x^3\,\mathrm{erg}$.

\subsection*{Analysis of Radial Oscillations}
\begin{figure}[h!]
\centering
\includegraphics[width=0.495\textwidth]{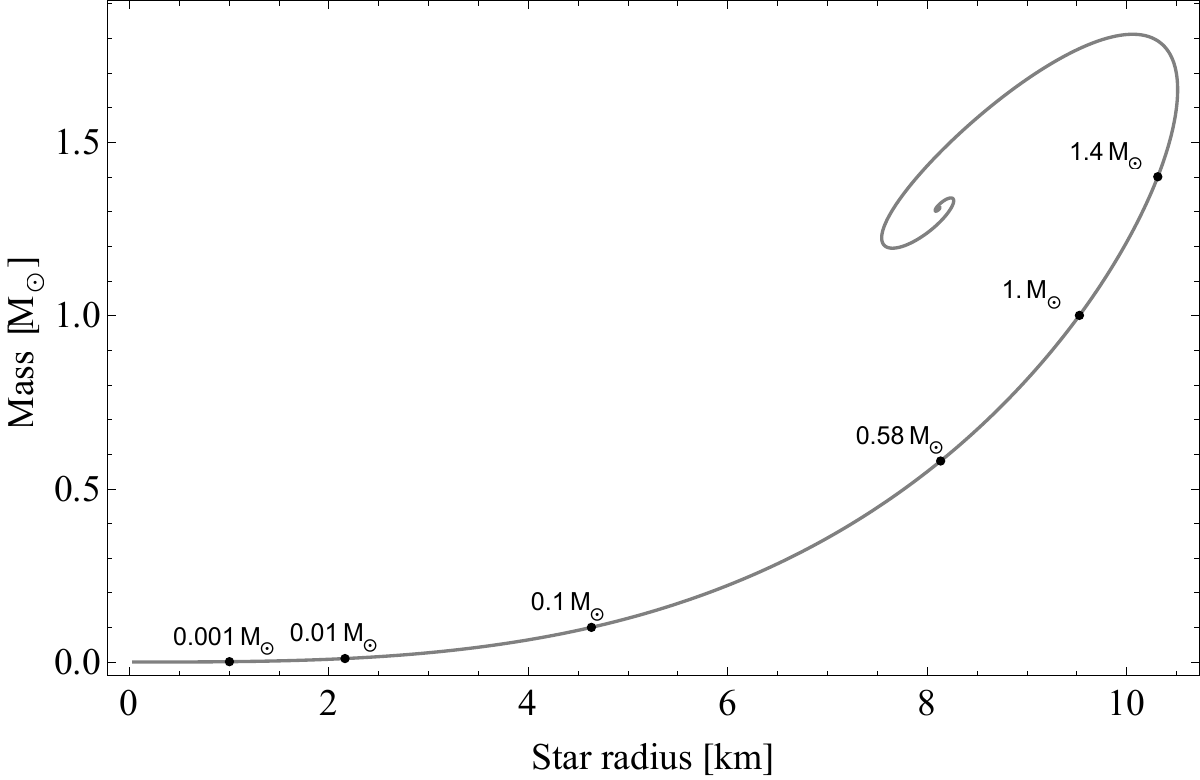}
\includegraphics[width=0.495\textwidth]{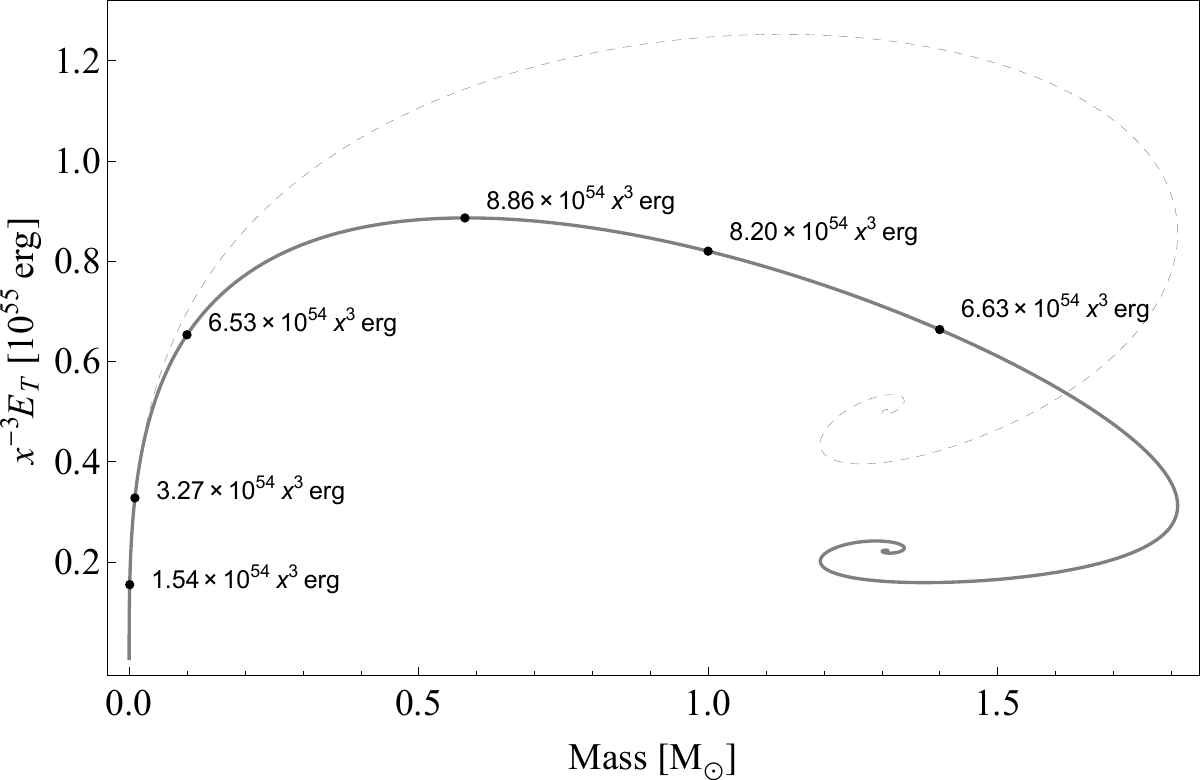}
\caption{\label{fig:rmerad}\small \textit{Left:} Mass-radius relation for \sqm ($B=60\,\mathrm{MeV\,fm}^{-3}$, $m_s=150\,\mathrm{MeV}$). \textit{Right:} Energy \eqref{eq:eradB} released by a monopole oscillation of amplitude $x$, ignoring or including a relativistic factor $H_{\Phi}$.}
\end{figure}

Small perturbations around equilibrium reveal that expanding the star lowers near-surface densities below \(n_0\), exciting \sqm to higher-energy states. Figure~\ref{fig:rmerad} shows the mass-radius curve for \sqm and the energy release \(\erad\) from a thin surface zone if $x$ is nonzero.

\subsection*{Conversion to Electromagnetic Energy}

Below saturation density, \sqm rapidly decays into photons, hadrons, leptons. We estimate the electromagnetic release rate \(P=E_{\text{exc}}/\tau\), with $\tau$ a few oscillation periods. Even $x=0.001$ yields $\sim1\,\mathrm{keV}$ per baryon near the surface, quickly emitted on a timescale of $\sim1\,\mathrm{ms}$.

\subsubsection*{Rapid Energy Release}
High amplitudes can produce multi-MeV per baryon, totaling $\sim10^{51}\,\mathrm{erg}$ at rates up to $10^{55}\,\mathrm{erg\,s}^{-1}$. Such violent events could fragment or dissolve the star altogether.

\subsection*{Baryon Density and Energy per Baryon in Excited SQM}
\begin{figure}[h!]
\centering
\includegraphics[width=0.495\textwidth]{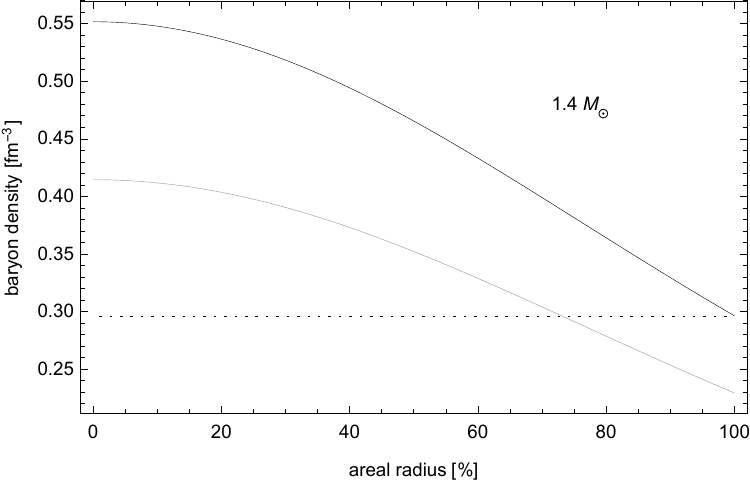}
\includegraphics[width=0.495\textwidth]{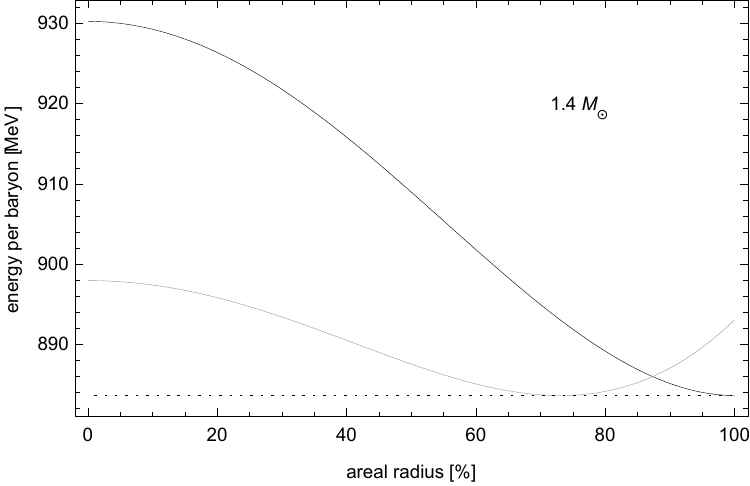}
\caption{\label{fig:star_density}Baryon density (left) and energy per baryon (right) for a $1.4\,\msun$ \sqm star in equilibrium (black) vs.\ a 10\% radial expansion (gray). Dashed lines show $n_0$ and its corresponding energy.}
\end{figure}

Figure~\ref{fig:star_density} illustrates how a $1.4\,\msun$ star’s surface region becomes unstable at $10\%$ expansion. The dashed line indicates $n_0$ saturation. The region $n<n_0$ gains $\epsilon_{\text{exc}}$ and rapidly radiates away.

\begin{table*}[thb!!!]
\setlength{\tabcolsep}{5pt} 
\centering
\begin{small}
\begin{tabular}{|c|c|c|c|c|c|c|c|}
\hline
Mass & $\nb$ & $R$ & $\ampl$ & $\erad$ & $r_c$ & $\nb(r{>}r_c)$ & $\meanexcen \left.(\excen\right|_R)$ \\
\hline
\hline
$1.0\,\mearth$ & $3.790{\times}10^{51}$ & $145.1\,\m$ & $0.001$ & $7.315{\times}10^{42}\,\erg$ & $0.0\,\m$ &  $3.790{\times}10^{51}$ & $1.205(1.221)\,\keV$\\
\hline
 &   &   & $0.001$ & $6.630{\times}10^{45}\,\erg$ & $10.29\,\km$ & $1.288{\times}10^{55}$ & $321.3(961.1)\,\eV$\\
 \cline{4-8}
$1.4\,\msun$ & $2.032{\times}10^{57}$ & $10.31\,\km$ & $0.01$ & $6.592{\times}10^{48}\,\erg$ & $10.06\,\km$ & $1.278{\times}10^{56}$ & $32.19(96.14)\,\keV$\\
 &  &  & $0.1$ & $6.160{\times}10^{51}\,\erg$ & $7.535\,\km$ & $1.158{\times}10^{57}$ & $3.321(9.450)\,\MeV$\\
 \hline
\end{tabular}
\end{small}
\caption{\label{tab:sqm_properties}\small 
Sample \sqm objects at different radial amplitudes. $\nb$ = total baryon number, $R$ = (areal) radius, $\ampl$ = fractional amplitude, 
$\erad$ = eventual radiated energy, $r_c$ = conversion radius above which $n<n_0$, $\meanexcen$ = average excitation per baryon, $\left.\excen\right|_R$ = excitation at the surface.}
\end{table*}
\FloatBarrier
\subsection*{Total Energy to Be Radiated}
For constant-density (planetary) \sqm, \eqref{eq:eradA} implies $\erad\propto x^2$. For larger \sqm stars, \eqref{eq:eradB} yields $\erad\propto x^3$. Table~\ref{tab:sqm_properties} summarizes energies, radii, and baryon numbers for sample objects, showing that even $x=0.001$ can release sizable energy.

\section{Discussion and Astrophysical Implications}
\label{sec:discussion}

\textbf{Sensitivity to Oscillations.}  
Our results highlight how \sqm surfaces respond acutely to radial oscillations. Even $x\sim0.1\%$ triggers noticeable energy release near $n_0$, quickly radiated as electromagnetic energy on $\sim1\,\mathrm{ms}$ timescales. Larger $x$ can unbind outer layers, fragmenting or dissolving the star.

\textbf{Implications for Gravitational Waves.}  
Energy lost to electromagnetic radiation changes the dynamics in \sqm binaries. Such systems might produce distinct gravitational wave signatures if partial dissolution occurs. Accounting for this effect in wave templates could refine interpretation of signals involving \sqm objects.

\textbf{Observational Prospects.}  
The short, intense photon bursts or transients could coincide with \sqm star oscillations or mergers. If fragmentation forms smaller \sqm bodies, these might persist as strange nuggets or be ejected.

\section{Conclusion and Future Work}
\label{sec:conclusion}

\noindent \textbf{Energy Conversion and Stability:}  
Even small radial amplitudes (\(x\lesssim0.001\)) can lead to $\sim1\,\mathrm{keV}$ per baryon above the surface, released in milliseconds. Larger amplitudes ($x\gtrsim0.01$) may exceed outer-layer binding energies, causing fragmentation or dissolution.

\noindent \textbf{Gravitational Wave Sources:}  
Damping by electromagnetic energy release can alter orbital evolution and waveforms in binary systems containing \sqm. More precise modeling could improve gravitational-wave detection and parameter extraction from such sources.

\noindent \textbf{Astrophysical Impact:}  
Oscillation-induced instabilities or fragmentation can shape the fate of \sqm stars, produce exotic transients, and generate smaller \sqm lumps. Observing these phenomena would bolster the case for SQM’s existence.

\noindent \textbf{Future Work:}  
\begin{itemize}
\item Investigate detailed radiation mechanisms and multiwavelength signatures of decaying \sqm surfaces.
\item Explore long-term evolution and repeated oscillation cycles in binary or cluster environments.
\item Incorporate SQM oscillation effects into gravitational-wave templates to test observational data for possible \sqm signals.
\end{itemize}

\bibliographystyle{JHEP}
\bibliography{2025LBJJ_oscillations_pos}

\providecommand{\href}[2]{#2}\begingroup\raggedright\begin{thebibliography}{10}

\bibitem{1984PhRvD..30..272W}
E.~{Witten}, \emph{{Cosmic separation of phases}},
  \href{https://doi.org/10.1103/PhysRevD.30.272}{\emph{\prd} {\bfseries 30}
  (1984) 272}.

\bibitem{1986ApJ...310..261A}
C.~{Alcock}, E.~{Farhi} and A.~{Olinto}, \emph{{Strange Stars}},
  \href{https://doi.org/10.1086/164679}{\emph{\apj} {\bfseries 310} (1986)
  261}.

\bibitem{1996astro.ph..4035W}
F.~{Weber}, C.~{Schaab}, M.K.~{Weigel} and N.K.~{Glendenning}, \emph{{Quark
  Matter, Massive Stars and Strange Planets}}, {\emph{arXiv e-prints}
  {\bfseries astro-ph/9604035} (1996) }
  [\href{https://arxiv.org/abs/astro-ph/9604035}{{\ttfamily
  astro-ph/9604035}}].

\bibitem{1984PhRvD..30.2379F}
E.~{Farhi} and R.L.~{Jaffe}, \emph{{Strange matter}},
  \href{https://doi.org/10.1103/PhysRevD.30.2379}{\emph{\prd} {\bfseries 30}
  (1984) 2379}.

\bibitem{2017ApJ...848..115H}
Y.F.~{Huang} and Y.B.~{Yu}, \emph{{Searching for Strange Quark Matter Objects
  in Exoplanets}}, \href{https://doi.org/10.3847/1538-4357/aa8b63}{\emph{\apj}
  {\bfseries 848} (2017) 115}.

\bibitem{Kuerban_2020}
A.~Kuerban, J.-J.~Geng, Y.-F.~Huang, H.-S.~Zong and H.~Gong, \emph{Close-in
  exoplanets as candidates for strange quark matter objects},
  \href{https://doi.org/10.3847/1538-4357/ab698b}{\emph{The Astrophysical
  Journal} {\bfseries 890} (2020) 41}.

\bibitem{2015ApJ...804...21G}
J.J.~{Geng}, Y.F.~{Huang} and T.~{Lu}, \emph{{Coalescence of Strange-quark
  Planets with Strange Stars: a New Kind of Source for Gravitational Wave
  Bursts}}, \href{https://doi.org/10.1088/0004-637X/804/1/21}{\emph{\apj}
  {\bfseries 804} (2015) 21}.

\bibitem{Kutschera_2020}
M.~Kutschera, J.~Ja{\l}ocha, {\L}.~Bratek, S.~Kubis and T.~K\k{e}dziorek,
  \emph{Oscillating strange quark matter objects excited in stellar systems},
  \href{https://doi.org/10.3847/1538-4357/ab97ab}{\emph{The Astrophysical
  Journal} {\bfseries 897} (2020) 168}.

\bibitem{PhysRevD.9.3471}
A.~Chodos, R.L.~Jaffe, K.~Johnson, C.B.~Thorn and V.F.~Weisskopf, \emph{New
  extended model of hadrons},
  \href{https://doi.org/10.1103/PhysRevD.9.3471}{\emph{Phys. Rev. D} {\bfseries
  9} (1974) 3471}.

\bibitem{bratek2024}
{\L}.~Bratek, J.~Ja\l{}ocha and M.~Kutschera, \emph{Phenomenological scaling
  relations for strange quark matter stars with a massive $s$ quark in
  gravitationally strong magnetic fields under the spherical symmetry
  approximation},
  \href{https://doi.org/10.1103/PhysRevD.110.083041}{\emph{Phys. Rev. D}
  {\bfseries 110} (2024) 083041}.

\end{thebibliography}\endgroup
\end{document}